# Energy Efficient Extreme MIMO: Design Goals and Directions

Stefan Wesemann, Jinfeng Du, *Member, IEEE*, and Harish Viswanathan, *Fellow, IEEE*

*Abstract*— Ever since the invention of Bell Laboratories Layer Space-Time (BLAST) in mid 1990s, the focus of MIMO research and development has been largely on pushing the limit of spectral efficiency. While massive MIMO technologies laid the foundation of high spectrum efficiency in 5G and beyond, the challenge remains in improving energy efficiency given the increasing complexity of the associated radio systems. With the substantial negative implications of climate change looming ever closer, minimizing energy use is a key dimension of achieving sustainability and is of paramount importance for any future technology. Thus, every aspect of future extreme MIMO system design, implementation, and operation will be scrutinized to maximize energy efficiency. An analysis of the massive MIMO 5G radio energy consumption at different loads leads to qualitative energy efficiency design goals for emerging extreme MIMO systems. Following this, we focus on novel operational and component technology innovations to minimize energy consumption.

## I. INTRODUCTION

WORLDWIDE numerous companies are racing to achieve carbon neutrality and have established carbon footprint reduction goals. Communications service providers (CSPs) are no exception. To reach that goal, the major focus will be on those aspects of operations that consume the largest amount of energy. Fig. 1 shows the energy consumption for a typical CSP, based on a survey [1] in 2021 of over 30 networks with a mixture of multiple generations and multiple radio access technologies across different domains of operations. It becomes clear that most of the energy consumption for a CSP is in network operations, among which the radio access network (RAN) infrastructure consumes the largest amount of energy. With the current mix of 3G/4G/5G, the radio unit (RU) contributes about 40% of the RAN energy consumption, comparable with the share of air conditioning at base transceiver station (BTS). With the ongoing rollout of the larger massive MIMO (mMIMO) RUs and the corresponding shift of processing functionality into the RU, the share of RU energy consumption will continue to grow in coming years [3],[4]. In response to that, 3GPP initiated a study on network energy savings [2], which includes the definition of an enhanced BTS energy consumption model, the introduction of energy consumption as a key performance indicator (KPI), as well as the study of techniques for improved energy savings at the BTS and user equipment (UE) side. A comprehensive survey on potential energy efficiency features and power consumption models can be found in [3]. Of course, it is not possible to simultaneously achieve peak performance across all metrics which is why the 3GPP study prioritizes idle/empty and low/medium load scenarios that allow for a flexibly tradeoff between the energy consumption and the other traditional metrics such as peak data rate, capacity, spectral efficiency, latency, and reliability.

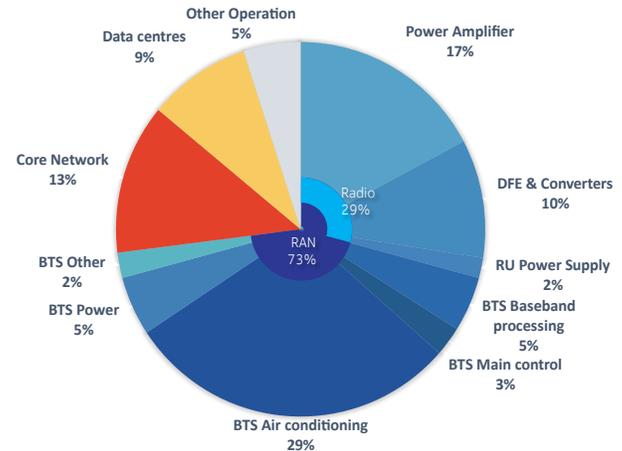

Fig. 1. Distribution of energy consumption for CSP operations, based on data from [1].

The purpose of this paper is to identify and expose key operational and component technology innovations that enable the evolution towards the next generation MIMO systems with extremely large arrays (over 1000 antenna elements), envisioned in new mid band (7 to 20 GHz) spectrum. In Section II, we begin with a brief look at the power consumption in today's mMIMO RUs to understand the potential areas where the next generation MIMO systems could be designed better. We also discuss what a good target for future RUs should be in comparison to 5G. In Section III, we describe the required innovations with a focus on energy efficiency (EE) aware scheduling, component deactivation, and energy saving concepts for the power amplifier (PA). In Section IV, we quantify the potential energy savings through a high-level model, and finally highlight research areas to further improve network energy efficiency in Section V.

Stefan Wesemann is with Nokia Bell Labs, 70469 Stuttgart, Germany (e-mail: stefan.wesemann@nokia-bell-labs.com).

Jinfeng Du and Harish Viswanathan are with Nokia Bell Labs, Murray Hill, NJ 07974, USA (e-mail: jinfeng.du@nokia-bell-labs.com, harish.viswanathan@nokia-bell-labs.com).

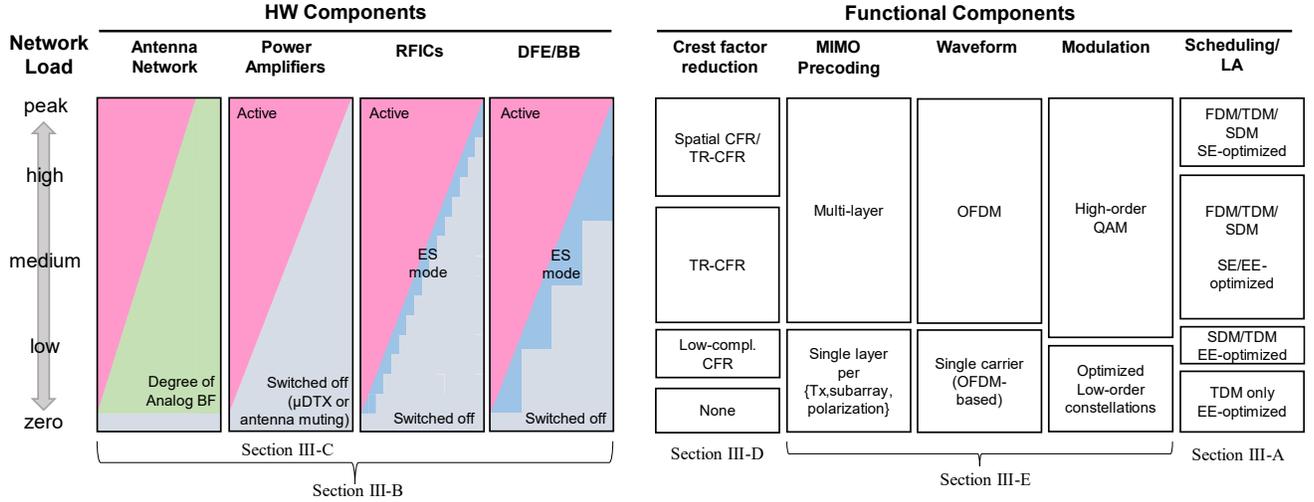

Fig. 2. Overview on EE concepts offered by the RU's hardware (left) and functional components (right). The left half shows for decreasing network load (from top to bottom), how antenna muting gradually sends the active hardware components (red) into energy saving (ES) mode (blue) and eventually switches those off (gray). The number of transceiver chains, implemented into a single RFIC (resp. DFE/BB chip), determines the granularity at which the components can be switched off. The right half illustrates the relationships between the functional components as a function of the network load.

## II. ENERGY EFFICIENCY GOALS FOR EXTREME MIMO

The peak capacity of a 64 transceiver (TRX) 5G mMIMO cell with 100 MHz bandwidth is about 20 times that of an LTE 20 MHz frequency division duplex cell [3] and is 90% more energy efficient in terms of watts per bits/sec. In other words, 5G consumes 10 times less power per bits/sec which translates to the LTE radio consuming 50% of 5G power at its peak capacity.

The RUs are, however, not always operated at their peak load because the RAN utilization is very uneven. It is estimated [5] that on average up to 70% of radio resources are idle, and 80% of sites only carries about 20% of the total traffic. Moreover, the load varies substantially over the daily 24-hour period, with peak load only occurring for a few hours.

Power consumption of today's RUs does not diminish automatically with reduced load. Lower loads imply a lower PA output power, which does not fully translate into a reduced PA power consumption because the PAs are operated in the high input (power) backoff regime where their efficiency is very poor. One commonly implemented feature to improve the PAs' energy consumption in low load regimes is micro-discontinuous transmission (μDTX), where sparse data traffic is grouped into high power bursts with low backoff, and thereby creating idle transmission periods at which the analog components are pushed into micro-sleep mode. In the current RUs, the digital processing power consumption does not decrease with the load because the digital frontend (DFE) and baseband (BB) processing platforms cannot be pushed into sleep mode. Hence the energy efficiency of currently deployed mMIMO systems is quite low as the load approaches zero. To bring power consumption close to zero, the cell essentially needs to be switched off. The time taken to become operational from this state is significant, hence it cannot be used for a temporary break in traffic. Clearly, this indicates the need for operational and technological innovations to further improve the energy efficiency at any load.

We expect future extreme MIMO (xMIMO) RUs to be designed for a peak capacity of about 10 times that of current mMIMO RUs to meet the growing demand. This can be realized using wider bandwidth (e.g., 400 MHz) and larger array sizes (e.g., 1024 antenna elements) in new mid band spectrum, expected in the range 7 to 20 GHz. The peak capacity should be achieved with at most twice the peak power consumption of current mMIMO RUs to ensure a manageable power supply and a higher EE (in terms of GB/kWh). To reach that goal, xMIMO RUs could leverage hybrid beamforming architectures in which an antenna network connects e.g., 128 or more TRX chains to the 1024 antenna elements.

However, it is even more important to design the new RUs so that the energy consumption scales better with the load, so that we nearly achieve "zero energy zero load" operation. The next section describes the key concepts that are expected to help in achieving this goal.

## III. OPERATIONAL AND TECHNOLOGICAL INNOVATIONS

This Section discusses operational innovations and component technology advancements that are expected to improve the energy efficiency of the future xMIMO RUs. The focus lies on the RU because with the introduction of mMIMO, many parts of the physical layer (PHY) signal processing (L1-Low, beamforming, DFE) have been shifted into the RU to keep the required fronthaul bandwidth small. The power consumption contributions from the distributed/centralized units corresponds to roughly 5 % of a mMIMO RU's full load power consumption (cf. measured power consumptions in [4]).

### A. Energy Efficient Scheduling and Link Adaptation

As illustrated in Fig. 2, the choice of the employed EE concepts will depend on the network load and thus needs to be under the control of an EE-aware scheduler. Today's scheduling and link adaptation (LA) strategies aim at maximizing the spectral efficiency (SE) by using MIMO OFDM with high constellation order and channel coding rate. While this is favorable at peak load, it does not provide the best tradeoff between SE and EE at medium to high load even in combination with μDTX. An EE-aware scheduling/LA approach, which selects the best combination of EE functionalities offered by the PHY, can provide significant EE gains particularly in the medium to high load regime. As shown in [6], reducing the average transmit (Tx) power is preferable over μDTX because the SE is a logarithmic function of the Tx power and linear with respect to the transmission time. Therefore, instead of transmitting short bursts each with high spectral efficiency, stretching data transmissions over multiple slots allows a linear reduction in SE (i.e., reduced constellation size and code rate) but an exponential reduction in the required total radiated power (TRP). Of course, reducing the TRP by increasing the PAs' backoff yields only diminishing energy savings. A more efficient way to reduce the TRP is antenna muting, as discussed in the next subsection.

The maximum possible TRP reduction is limited by the QoS constraints such as minimum data rate or latency budget. Another limiting factor for TRP reduction is the need for a sufficiently high SNR at the UE side to ensure a proper channel estimation.

In addition to the TRP, further parameters such as MIMO mode, waveform & modulation, crest factor reduction mode, antenna muting, and beamforming need to be controlled by scheduler/LA, which calls for further research such as [7],[8].

### B. Antenna Muting

Whenever possible, the EE-aware scheduler/LA reduces the spectral efficiency and thus allows for a TRP reduction particularly at medium to high network load. With today's PAs, a TRP reduction is realized by increasing the backoff, which translates only partially into a PA power saving as the PA's line-up power added efficiency (PAE) rapidly drops with the backoff. An alternative to an increased backoff is the antenna muting where the scheduler/LA determines the number of active Tx chains, depending on the required TRP and array gain. The remaining set of active PAs still operates in their optimized low backoff regime. Further, parts of the analog and digital front end (AFE & DFE) can be sent into sleep mode or even deactivated (cf. Fig. 2) to save additional power, as the antenna muting periods are expected to be longer than those of μDTX. One should note that EE-aware scheduling/LA with antenna muting can already be pushed into today's mMIMO RUs via software update, as the μDTX functionality from the AFE is reused. In future radio platforms, the muting capabilities should be extended towards the DFE and BB processing to increase the power savings.

A key challenge with antenna muting is the loss in array gain which affects the coverage. This can be alleviated by adding flexibility into the antenna network, so that the RU can dynamically adjust the amount of analog beamforming needed to maintain the array aperture and thus array gain. The increased cost and size will determine the viability of this approach. In addition, research is needed on coverage enhancement features (e.g., repeated transmissions) for broadcast, synchronization, and/or control channels.

Another limitation that arises from muting larger numbers of Tx chains is the restriction on the supported number of MIMO layers. Further research is needed to determine the network load conditions under which an increased backoff or μDTX is preferrable over antenna muting.

### C. Advanced PA Technologies and Operation Regimes

An important research direction is improving the PA's line-up PAE which dictates its power consumption. Especially at high network loads where the full TRP is needed, the PAs' power consumption is dominating. State of the art (GaN-based) PAs offer ~40% line-up PAE at 7 dB backoff. A promising new solution is the Digital Doherty approach from [9] which can provide a higher line-up PAE (e.g., > 50% at 6.6 dB backoff) because it does not need an input impedance matching and eliminates quiescent currents in the output stages. Another concept for higher PAE is the Polar Tx [10] which offers 52% PAE at 7.1 dB backoff. The challenge with both concepts is the increased manufacturing cost which stems from the complicated flip-chip technology needed by the Digital Doherty, or the need for "calibrated" silicon to ensure a proper matching of the amplitude and phase branches in the Polar Tx.

A second PA-related research direction is the reduction of the DPD complexity. GaN-based PAs are highly non-linear and require an online linearization to ensure good signal quality (in terms of error vector magnitude, EVM) and limited out of band emissions (measured e.g., by the adjacent channel leakage ratio, ACLR). In today's radios, the DPD module's power consumption constitutes the largest portion in the DFE. Therefore, easier-to-linearize PAs that admit a low-complexity DPD are desirable. The Digital Doherty PA and Polar Tx can help in this regard. An alternative relaxation for the DPD requirements and thus complexity can arise from the use of advanced non-linear (e.g., machine learning based) receivers [11] which can handle the in-band distortions caused by operating the PA in the nonlinear regime (i.e., with very small backoff). The task of the DPD would then resort to the control of the out-of-band emissions.

### D. Advanced Crest Factor Reduction

The crest factor reduction (CFR) module reduces the peak-to-average-power-ratio (PAPR) of the Tx signals by clipping the signal peaks while keeping the in-band signal distortions (i.e., EVM), the out-of-band emissions (e.g., ACLR) and the maximum TRP under control. To translate the reduced PAPR into a lower PA power consumption, either the PA is re-designed for a lower peak power, or concepts such as drain voltage modulation are added. For class-B PAs, this would yield an improved PAE and thus lower power consumption at all loads, besides a reduced PA silicon size and cost. For Doherty PAs, a reduced power consumption is only achievable at higher backoff where the auxiliary PA is inactive. Without a PA re-design, the main benefit from an advanced CFR is the improved EVM (e.g., 1%) at maximum TRP, as needed at peak network load.

Today's CFR modules are based on the clipping & filtering method, operate separately on the individual Tx chains, and guarantee 7.5 dB PAPR with 3.5% EVM and <-50 dB ACLR. For multi-antenna RUs, the spatial domain can be additionally exploited for dumping clipping noise, yielding the so-called spatial CFR which combines in its ultimate form the MIMO precoding and CFR operation. As shown in [12], PAPRs in the order of 5 to 6 dB become achievable. The underlying challenges are an increased computational complexity, the need for a monolithic CFR processing jointly across all Tx chains, and the presence of accurate channel state information to keep the spatial directions towards the UEs clear from perturbations. Further, at aggressive PAPR targets the amount of additionally radiated perturbation power can become significantly large [13] and violate the maximum TRP limit.

At low and medium network loads, enhanced PAPR reductions are possible with the much simpler (i.e., per-antenna processing) tone-reservation (TR) based CFR. In those operation regimes, the EE-aware scheduler can free up (inband) frequency domain resources that are used by the CFR to dump larger amounts of clipping noise. Further, the lowered spectral efficiency implies a larger EVM budget which admits a higher amount of in-band distortions on the data carriers. Here, again, the Tx signal power loss due to radiated perturbations needs to be carefully controlled. Towards 6G, concepts such as active constellation shaping/extension as well as partial transmit sequences (esp. for single carrier waveforms) are potential candidates for which the Tx signal power loss as well as spectral efficiency loss needs to be carefully assessed.

### E. Adaptive Waveforms & Modulations

The downlink Tx signal's PAPR grew over the past generations of cellular standards and reached its peak in 4G and 5G with OFDM which exhibits (without the CFR) a PAPR of 10 to 12 dB. This led to the implementation of complex CFR and DPD methods in the DFE. For the uplink, a single-carrier waveform (i.e., DFT-s-OFDM) has already been standardized in 4G uplink whose low-PAPR characteristic significantly improved the coverage and reduced the power consumption at the UE side. In 3GPP release 18 (i.e., 5G-Advanced), even a dynamic switching between OFDM and DFT-s-OFDM uplink waveforms is discussed.

For adaptive downlink waveforms, similar concepts are proposed for 3GPP release 19. Important for the 6G waveforms is the support for multi-RAT spectrum sharing (possibly at resource-element level), which is intended to alleviate the 5G to 6G migration. This, however, narrows down the scope to OFDM-based single carrier (SC) waveforms like DTS-s-OFDM and variations of it such as Known Tail (KT)-DFT-s-OFDM. The PAPR of the SC waveforms grows with modulation symbol alphabet size which is why a CFR is still needed when higher-order QAM is used. The CFR's computational complexity can be, however, significantly reduced because no iterative clipping & filtering is required, as demonstrated in [14]. Moreover, at low network loads where

QPSK modulation would be sufficient, one could anticipate even the absence of a CFR to save digital processing power. An interesting research direction are probabilistic and geometric constellation shaping schemes which could help to maintain the low PAPR characteristics also for higher order constellations. Also, active constellation shaping/extension concepts which are typically discussed in the CFR context constitute a promising direction too.

The key challenge of SC waveforms is MIMO precoding, which results in a weighted combination of multiple SC waveforms with increased PAPR and loss of advantage over OFDM when the number of MIMO layers is high. More research is needed to identify novel (non-linear) precoding schemes that preserve the low-PAPR characteristics. Without such solutions, the usage of the SC waveforms in the downlink would be limited to one layer per Tx chain which either results in an array gain loss or a spatial multiplexing loss, the latter only tolerable at low network loads. Similarly, SC waveforms lose their low-PAPR characteristics when multiple signals are multiplexed in the frequency domain, which would limit their usage to pure time division multiplexing (TDM). Finally, it needs to be studied whether the SC waveforms are also suitable for the transmission of the broadcast, synchronization, and control channels as well as reference signals.

### IV. ENERGY EFFICIENCY ASSESSMENT

In this section, we develop a high-level power consumption model for RU, building on top of the excellent work [2], [3],[15], to assess the energy saving potentials of selected EE concepts by evaluating the power consumption and the supported network load for different RU configurations.

### A. mMIMO and xMIMO RU Configurations

For the baseline configuration, we assume a mMIMO RU as deployed today, with 3.5 GHz carrier frequency and $M = 64$ transceiver chains, $L = 16$ spatial layers (cf. Table 1), operating in time-division duplex with an uplink-to-downlink ratio $\beta = 0.75$ (i.e., 75 % of the slots are used for downlink transmission). To cope with the varying network load, the power saving features µDTX and µDRX are used to deactivate the analog components such as PAs and RFICs (including drivers and LNAs). This mMIMO RU does not support the deactivation of the DFE and BB processing.

The xMIMO RU leverages a hybrid beamforming architecture, in which an antenna network (cf. insertion loss $\varepsilon$ in Table 1) connects $M = 128$ transceiver chains to 1024 antenna elements. The xMIMO RU operates at 7 GHz, has $b = 4$ times larger occupied bandwidth, and serves $L = 32$ layers simultaneously. The higher array gain and higher TRP help to compensate the 6 dB higher path loss ($f_c^2$) and the 6 dB lower power spectrum density from using 4 times wider bandwidth in downlink. For the increased building penetration loss, the 3GPP TR 38.901 low building penetration loss model implies about 1 dB additional loss. The xMIMO RU as anticipated in near future uses an EE-aware scheduling/LA method for TRP reduction through antenna muting. In addition to switching off PAs and RFICs, the RU implements an energy saving (ES) mode for the DFE/BB (e.g., clock gating for the processing pipes, deactivation of interfaces) and a chip-level deactivation of the DFE/BB processors, which switches off a chip as soon as all connected TRX chains are muted. Further, we anticipate $\eta_{DSP} = 2$ times enhancements in the compute intensity per Watt of the digital processing platforms, and we expect improvements in the PA's line-up efficiency $\eta_{PA}$ (cf. Table 1). It should be noted that EMF exposure mitigation methods may be needed for the xMIMO RU where sufficient compliance distance can't be maintained.

| RU/MIMO configuration | mMIMO | xMIMO |
|---|---|---|
| Carrier Frequency | 3.5 GHz | 7 GHz |
| Number of antenna elements | 192 | 1024 |
| Number of transceiver chains $M$ | 64 | 128 |
| Number of spatial layers $L$ | 16 | 32 |
| Total radiated power $P^{TRP}$ | 200 W | 400 W |
| Bandwidth (oBW) | 100 MHz | 400 MHz |
| Antenna network insertion loss $\varepsilon$ | 1.1 dB (filter & splitter loss) | 2.3 dB (1.6dB filter & splitter loss, 0.7dB phase shifter loss) |
| Layer SINR $\gamma$ | 25 dB | 21 dB |
| Technology/EE concepts | Today | Future |
| Network load adaptation | µDTX and µDRX | EE-aware sched./LA with antenna muting |
| PA's line-up PAE $\eta_{PA}$ | 45 % | 55 % (Digital Doherty PA) |
| Relative compute intensity per Watt $\eta_{DSP}$ | 1 | 2 |

Table 1. Model parameters for the mMIMO and xMIMO radio units and the RU technology/energy efficiency concepts.

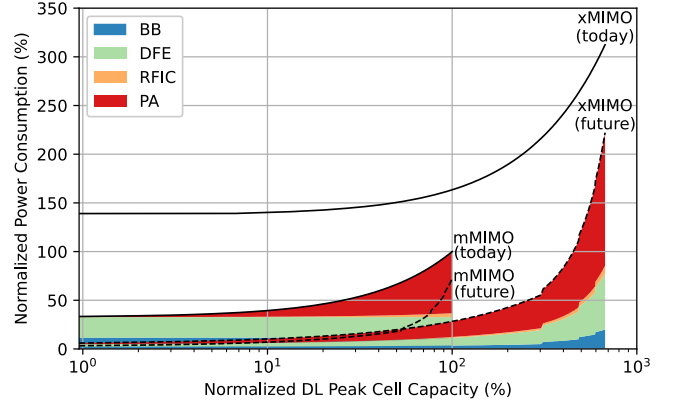

Fig. 3. Normalized RU power consumption as a function of the supported network load, where the maximum DL cell capacity and peak power consumption of today's mMIMO RU is used as baseline reference. In this comparison, today's mMIMO/xMIMO estimates leverage µDTX/DRX and the future mMIMO/xMIMO estimates (dashed lines) adopt EE-aware scheduling/LA, antenna muting, energy saving modes, chip-level deactivation, and advanced PAs.

For simplicity we do not impose latency constraints onto the data channels. Also, the (uplink) coverage loss due to antenna muting is assumed to be compensated by a flexible antenna network that maintains the array gain, and/or special 6G PHY features such as repeated transmissions.

## B. Power Consumption and Data Rate Models

Following the BTS power consumption model from 3GPP [2], we decompose the RU into its main four components: PA, RFIC, DFE and BB. The latter two components are assumed to be implemented on $N_{chip} = 4$ processors.

The full load power consumption of the PA depends on the following parameters (cf. Table 1): nominal TRP $P^{TRP}$, antenna network insertion loss $\varepsilon$, PAs' power supply efficiency $\eta_{PA}^{PSU} = 0.94$, and PAs' line-up efficiency $\eta_{PA}$. One should note that the modelled transmission strategies do not realize the TRP reduction via increased backoff. Hence, the PAs are always operated in their optimized high PAE regime.

The full load power consumption per TRX chain (dominated by RFIC, DFE and BB/L1-low components) is given by
- TX: $\hat{P}_{RFIC}^{driver} = 0.5\,W$, $\hat{P}_{DFE/BB}^{Tx} = 1.75\,W$,
- RX: $\hat{P}_{LNA}^{RX} = 0.3\,W$, $\hat{P}_{DFE/BB}^{Rx} = 1.5\,W$.

The average power consumption of these four components varies with the network load. To keep the model simple, we assume that the network load variation is the same in uplink and downlink. Further, we ignore the power consumptions caused by the transmission of synchronization & broadcast channels, reference signals, and due to sleep state transitions. For µDTX and µDRX, the average power consumption of the PAs and RFICs scales linearly with the fraction of active transmission and reception time, denoted by $t \in (0,1]$. Similarly for antenna muting, the power consumption scales linearly with the fraction of active transmit and receive chains, denoted by $m \in (0,1]$.

Moreover, the RU has a static (i.e., load independent) power consumption component $\hat{P}_{DFE/BB}^{STATIC} = 2W$ (per TRX chain) which arises in the DFE, BB and in various smaller system components even when the ES mode is used.

The power consumption for the described RU configurations can be calculated by

$$P = \beta M \left[ mt \left( \frac{\varepsilon}{M \eta_{PA}^{PSU} \eta_{PA}} P^{TRP} + \hat{P}_{RFIC}^{driver} \right) + \frac{mb}{\eta_{DSP}} \hat{P}_{DFE/BB}^{Tx} \right]$$
$$+ (1-\beta)M \left[ mt \hat{P}_{RFIC}^{LNA} + \frac{mb}{\eta_{DSP}} \hat{P}_{DFE/BB}^{Rx} \right] + \frac{M\lceil mN_{chip}\rceil}{\eta_{DSP}N_{chip}} \hat{P}_{DFE/BB}^{STATIC},$$
$$(1)$$

The first two terms in Eq. (1) correspond to the dynamic power consumptions at transmitter and receiver, respectively, that vary with the use of µDTX, µDRX and antenna muting. The third term remains static for today's mMIMO RUs, while future xMIMO RUs can reduce it by a step-wise deactivation of $N_{chip} - 1$ chips.

To quantify the supported network load without running complex system level simulations, we adopt the Shannon formula to estimate the RU's (peak) downlink rate by

$$R = \beta \cdot b \cdot t \cdot l(m) \cdot log_2 \left(1 + m^2 \cdot \frac{L}{l(m)} \cdot \gamma\right),$$
$$(2)$$

where $\gamma$ denotes the per-layer signal-to-interference-plus-noise ratio (SINR). For simplicity, we assume the same SINR for all MIMO layers, as tabulated in Table 1. For antenna muting, we include the function $l(m) = \min(L, m \cdot M)$ which accounts for the reduction in the number of spatial layers when reaching very small numbers of active transmit chains (i.e., less transmit chains than $L$ layers).

*C. Power Consumption Analysis*

Fig. 3 shows the breakdown of the relative power consumption for today's mMIMO and future xMIMO RU configurations, using today's mMIMO peak cell capacity as 100% reference for the normalized DL cell capacity and its peak power consumption as 100% reference for the normalized power consumption. To isolate the effects of more antennas/higher array gain and advanced EE concepts, we also include the power consumption curves for a future mMIMO RU and an xMIMO RU as if build today. The curves are obtained by sweeping through the parameter $t \in (0,1]$ for today's RUs, and $m \in (0,1]$ for the future RUs. As shown in the figure, the xMIMO configuration provides 6.7x higher peak cell capacity resulting from 2 times more spatial layers, 4 times larger bandwidth, and 4 dB lower layer-SINR. With advanced EE features, the xMIMO RU's power consumption at peak load will be roughly twice as much with respect to today's mMIMO configuration, and it would be 3x if the xMIMO RUs were build today without the advanced EE features. To compute a daily average for the mMIMO's (resp. xMIMO's) power consumption, we assume 50% load for 8 hours, 30% load for 10 hours, and 5% load for 6 hours, where 100% load corresponds to mMIMO's (resp. xMIMO's) peak cell capacity. Our model indicates a daily average power consumption for the future xMIMO RU will be 17% less than the daily average of today's mMIMO baseline. For comparison, without advanced technology/EE concepts, the daily average power of xMIMO would be 3.6x of mMIMO, which highlights the importance of advanced EE features to greatly improve RAN energy efficiency. Section V lists further EE concepts that could potentially help to achieve the goal of 50% energy consumption reduction for the xMIMO RUs.

One should note that our model captures only the data channel; that is, advanced energy saving concepts for the common/broadcast channels as well as signals (e.g., via reduced signaling) need to be leveraged for further power reduction, particularly for empty cells and low load conditions.

## V. DISCUSSION AND CONCLUSIONS

Energy efficiency will be an important metric for the next generation networks, at least as important as the traditional performance metrics such as throughput, latency, reliability and scalability. While 5G mMIMO improved over 4G substantially in terms of the bits/joule metric at peak load, it is highly inefficient during lightly loaded conditions. An important requirement for the next generation xMIMO design should be to ensure that energy consumption scales down gracefully with traffic, approaching zero energy at zero traffic. With more than a ten-fold capacity increase expected at peak loads, xMIMO RUs should leverage various technology advances to reduce overall average energy consumption by 50% and to improve the bits/joule metric by five times at peak load and by 20 times on average, as compared to 5G.

For reaching the 50% reduction goal, however, further energy saving concepts such as the flexible usage of hybrid beamforming in combination with antenna muting, compute-in-memory for efficient signal processing, and adaptive waveforms to offload the burden from CFR and DPD are urgently needed. Network densification in conjunction with or in addition to distributed/cell-free mMIMO could push further into energy efficient wireless transmissions if the site acquisition challenge could be addressed. AI/ML-assisted cell shutoff management as well as pooling gains from centralization may also help to further reduce the power consumption.

To achieve dramatic improvement in energy efficiency while providing higher capacities when required, the industry could consider the following new approaches in its next-generation standards:

- Adaptive waveforms to reduce PAPR while avoiding complex digital pre-distortion
- Frequency domain shaping or coding techniques to reduce PAPR by exploiting unused bandwidth
- Facilitate antenna muting by ensuring coverage solutions for different array sizes
- Mesh fronthaul to ensure resilience and increase pooling gains for centralized functions
- Exposure of energy usage statistics to consumer apps to promote energy awareness
- Adopt Total Life Cycle Assessment as it plays a key role in the sustainability and carbon footprint (energy consumed) of deploying a new network.

## ACKNOWLEDGMENT

The authors would like to thank Harri Holma, Preben Mogensen, Navin Hathiramani, Gagandeep Bhatti, Daniela Laselva, Susanna Kallio, and many others at Nokia for their helpful comments and suggestions.

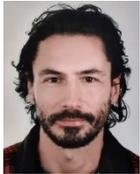

**Stefan Wesemann** is head of the Transceiver Research Department Germany & France in Nokia Bell Labs. He received the Dipl.-Ing. (M.Sc.) in Information Systems Engineering in 2006, and the Dr.-Ing. (PhD) in Electrical Engineering in 2014, both from Dresden University of Technology. Since joining Nokia Bell Labs in 2014, Stefan has worked on massive MIMO algorithms and system design and supported the productization of those. His research focus is on novel transceiver technologies, extreme & distributed MIMO design, and disruptive PHY concepts for improved energy efficiency. He is a Bell Labs Distinguished Member of Technical Staff.

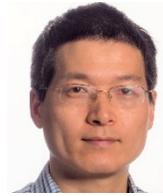

**Jinfeng Du** (Member, IEEE) is leading the Department of Radio Systems USA in Nokia Bell Labs. He received the B.Eng. degree in electronic information engineering from the University of Science and Technology of China (USTC), Hefei, China, in 2004, and the M.Sc., Tekn. Lic., and Ph.D. degrees from the Royal Institute of Technology (KTH), Stockholm, Sweden, in 2006, 2008, and 2012, respectively. He was a Post-Doctoral Researcher with the Massachusetts Institute of Technology (MIT), Cambridge, MA, USA, from 2013 to 2015. Since joining Bell Labs in 2015, he has focused on fundamentals and disruptive concepts for wireless communications, especially in communication theory, radio system design and optimization, millimeter-wave propagation measurements and modeling.

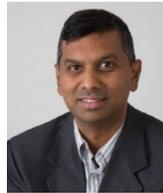

**Harish Viswanathan** (Fellow, IEEE) is Head of Radio Systems Research in Nokia Bell Labs. He received the B. Tech. degree from the Department of Electrical Engineering, Indian Institute of Technology, Chennai, India and the M.S. and Ph.D. degrees from the School of Electrical Engineering, Cornell University, Ithaca, NY. Since joining Bell Labs in October 1997, he has worked extensively on wireless research ranging from physical layer to network architecture and protocols including multiple antenna technology for cellular wireless networks, multi-hop relays, network optimization, network architecture, and IoT communications. From 2007 to 2015, Harish was in the Corp CTO organization, where as a CTO Partner he advised the Corporate CTO on Technology Strategy through in-depth analysis of emerging technology and market needs. He is a Fellow of the IEEE and a Bell Labs Fellow.